\documentclass{aa520}
\usepackage{graphicx}
\usepackage{times}
\usepackage{psfig}
\usepackage{amsmath}
\usepackage{natbib}
\bibpunct{(}{)}{;}{a}{}{,}
\newcommand{\sqdegr}{\raisebox{0.65ex}{\tiny\fbox{$ $}}\,$^{\circ}$}

\def\hide#1{}

\def\mpcoh{{\,h^{-1}\,\rm Mpc}}
\def\logten{\log_{10}}

\def\bj{B}

\begin{document}

\title{COMBO-17 measurements of the effect of environment on the type-dependent galaxy luminosity function} 
\author{S. Phleps\inst{1,2} \and  C. Wolf\inst{3} \and J.~A. Peacock\inst{1} \and
K. Meisenheimer\inst{4}\and E. van Kampen\inst{5}}
\offprints{S.~Phleps (sphleps@mpe.mpg.de)}

\institute{
Institute for Astronomy, University of Edinburgh, Royal Observatory,
Blackford Hill, Edinburgh EH9 3HJ, UK
\and Max-Planck-Institut f\"ur Extraterrestrische Physik,
Giessenbachstra{\ss}e, D-85748 Garching, Germany
\and Department of Physics, University of Oxford, Denys Wilkinson
Building, Keble Road, Oxford OX1 3RH, UK 
\and Max-Planck-Institut f\"ur Astronomie, K\"onigstuhl 17, D-69117
Heidelberg, Germany
\and Institut f\"ur Astrophysik, Leopold-Franzens-Universit\"at Innsbruck, Technikerstra{\ss}e 25, 6020 Innsbruck, Austria 
} 

\titlerunning{The dependence of the galaxy luminosity function on the local density contrast}
\date{}
\date{Received 1. 12. 2006 / Accepted 16. 03. 2007}

\abstract{We have developed a method to calculate overdensities in multicolour
surveys,  facilitating a direct comparison of the local density contrast
measured using galaxy samples that have different redshift error
distributions, i.e. for red and blue, or bright and faint galaxies,
respectively.  We calculate overdensities in small redshift slices
{\bf ($\Delta z =0.02$, which at $z=0.3$ corresponds roughly to $\Delta
r_{\rm comoving}=53 h^{-1}$\,Mpc) for $9176$ galaxies with $R\leq23.65$, 
$M_{B}({\rm Vega})-5\log h\leq-18$, and $z\leq 0.7$}, in three COMBO-17
fields {\bf (measuring
$31'\times31'$ each). The mean redshift errors of this sample are
approximately $\sigma_z/(1+z)\simeq 0.015$.} In the Chandra
Deep Field  South we identify a
region that is underdense by almost a factor 2 compared to the other two
fields in the same redshift range ($0.25\la z \la 0.4$). This can be
used for an investigation of the variation of the colour-dependent  luminosity
function with environment: We calculate the luminosity function in this
redshift range for red
sequence and blue cloud galaxies {\bf (as defined by
\citet{Bell04})} in each of the fields
separately. While the luminosity function of the blue galaxies remains
unaffected by different density contrasts, the luminosity function of
the red galaxies  clearly has a
more positive faint-end slope in the 
Chandra Deep Field South {\bf as compared to the other two
COMBO-17 fields}. The underdensity there is thus mainly
due to a deficiency of faint red 
galaxies. This result is in qualitative agreement with the trends seen at $z=0.1$,
e.g. in the 2dFGRS \citep{Croton05}{\bf , or in the SDSS \citep{Zandivarez06}}.
}
\maketitle
\section{Introduction}

It has been known for more than 25 years  that in a statistical sense galaxy properties
depend on the local environment: there is a clear
trend  for early-type systems to concentrate in high-density regions
(\citealp{Dressler80,Dressler97}). This dependence on
environment must hold important information about the history of galaxy formation,
so it is important to study the connection between the
properties of the galaxies and local galaxy density in greater detail.

The type-dependent luminosity function (the luminosity function calculated
for different galaxy types) directly quantifies how the colours and
luminosities of galaxies are influenced by their environment.
With the advent of deep redshift surveys, it has become feasible
to measure the luminosity function 
for field and cluster galaxies respectively
(e.g.
\citealp{Valotto97,Ratcliffe98,Trentham98,Marinoni99,Ramella99,TrenthamTully02,Martinez02,Christlein03,Zandivarez06});
eventually, statistical power reached the stage where the type-dependent luminosity
function could be estimated directly in regions of differing density contrast
(e.g. \citealp{Bromley98,Gray04}).

To date, the most comprehensive analysis of this type has been that of
\citet{Croton05}, who used  the 2dFGRS data \citep{Colless03} to investigate luminosity
functions split by galaxy colour in different environments. Since the 2dFGRS
galaxies have spectroscopic redshifts, it was feasible to calculate
the local density contrast in spheres of $8 h^{-1}$\,Mpc radius, and
since the number of galaxies and the observed dynamic range of
overdensities is large, the sample could be divided
into six bins of density contrast.

If the dependence of galaxy properties on environment reflects the
formation history of the galaxies, it is clearly of great interest to
carry out similar investigations at higher
redshifts. {\bf \citet{Ilbert06a} used the VIMOS-VLT Deep Survey (VVDS; see
\citealp{LeFevre05}) to measure the environmental
dependence of the total luminosity function in a sample of 6582 galaxies
with spectroscopic redshifts below $z<1.5$. They also 
investigated the galaxy luminosity function per morphological type up
to $z=1.2$ \citep{Ilbert06b}, per spectral type up to $z=1.5$
\citep{Zucca06}, and the build-up of the colour-density
relation \citep{Cucciati06}}.

\citet{Cooper06a} used a larger sample of 19,464 DEEP2 \citep{Davis03}
galaxies with {\bf spectroscopic} redshifts in the  range $0.4 \leq z \leq 1.35$. They measured
the evolution of the colour-density relation and found the fraction of
red galaxies to depend strongly on environment out to $z\simeq 1$.
They also investigated the type-dependent luminosity function, and
found a good general agreement between their measurement and the
COMBO-17 data from \citet{COMBOMain04}, but did not split the sample
by environment (\citealp{Faber06,Willmer06}), which is explored in
the present paper.

{\bf Compared to spectroscopic redshift surveys, multicolour photometric
redshift surveys have generally larger number statistics and
completeness to greater depths}, and hence 
can be used to measure the dependence of the luminosity function
on environment at intermediate redshift ($z\la 1.2$) in greater
statistical detail -- provided the influence of the
redshift inaccuracy on the measurement of the overdensities is well
understood. 
Using the COMBO-17 survey \citep{COMBOMain04}, we will
demonstrate in this paper how local overdensities can be 
computed and how the influence of the redshift errors can be
treated. We will then show how the dependence of the (Schechter) luminosity
function on the environment can
be investigated, and present the results.

This paper is structured as follows: The dataset  used in our study
(COMBO-17) is introduced in Section \ref{COMBO}.   
In Section~\ref{Method} we 
describe the influence of redshift inaccuracies on the measurement of the
local density contrast using a COMBO-17 mock catalogue, which is based on an $N$-body
simulation combined with a semi-analytic model for galaxy evolution
(\citealp{vanKampen99,vanKampen05}). Following this
exercise, we calculate the
local density contrast in three COMBO-17 fields and measure the
luminosity function for red sequence and blue cloud galaxies {\bf (as
defined in \citealp{Bell04})} in the
redshift range $0.25\leq z \leq 0.4$ (see 
Section \ref{Results}). The results are discussed in Section \ref{discussion},
and a brief summary and an outlook is given in
Section \ref{Summary}.

We assume a flat cosmological geometry 
with $\Omega_m=0.25$; all lengths quoted are in comoving units, and $h=H_0 / 100 \rm
\,km\,s^{-1}\,Mpc^{-1}$. All magnitudes are quoted with a Vega zero point.

\section{Database: The COMBO-17 Survey}\label{COMBO}
To date, COMBO-17 ({\bf C}lassifying {\bf O}bjects with {\bf M}edium
{\bf B}and {\bf O}bservations in 17 filters) has surveyed three
disjoint $\sim 31'\times 30'$ 
southern equatorial fields (the 
Chandra Deep Field South [CDFS], A901 and S11 field,
respectively; for their
coordinates see \citealp{Wolf03}, W03 in the following) 
to deep limits in $5$ broad and $12$ 
medium passbands, covering wavelengths from $400$ to $930$\,nm. {\bf The
classification is reliable for $R\la 24$}. A
detailed description of the survey along with filter
curves can be found in \citet{COMBOMain04}.

Galaxies were detected on the deep $R$-band images by using {\bf SE}xtractor
\citep{Bertin96}. The spectral energy distributions (SEDs) for $R$-band detected objects
were measured by performing seeing-adaptive, weighted-aperture
photometry in all $17$ frames at the position of the $R$-band detected
object. 

\subsection{Photometric redshifts}

Using the 17-band photometry, objects are classified  using a scheme
based on template spectral energy distributions (SEDs)
\citep{Wolf01a,Wolf01b}. Each
object is also assigned a redshift (if it is
not classified as a star).  The  redshift errors in this
process depend on the magnitude and type of the object. The galaxy
redshift estimate quality has been tested by comparison 
with spectroscopic redshifts for almost 1000 objects (see \citealp{COMBOMain04}).
At bright limits $R < 20$, {\bf the redshift errors are approximately}
$\epsilon_z/(1 + z) \simeq 0.01$, and the error is dominated by mismatches between
template and real galaxy spectra. This error can contain a systematic component
that is dictated by the exact filter placement, but these `redshift focusing'
effects are of the order of magnitude of the random redshift errors for $z<1$ and are 
unimportant for the current analysis.
At the median apparent magnitude
$R \simeq 23$, $\epsilon_z/(1 + z) \sim 0.02$. For the faintest
galaxies, the redshift accuracy 
approaches those achievable using traditional broadband photometric
surveys, $\epsilon_z/(1 + z)  \ga 0.05$. We thus restricted our analysis
to galaxies with $R< 23.65$.  
In order to define a volume limited sample at $z\le 0.7$, we furthermore select
galaxies with restframe $B$-band magnitudes $M_B - 5 \log_{10}h
\leq-18.0$, and find $9176$ galaxies with $z\leq 0.7$ that fulfil these requirements.
There is no point in trying to correct for the incompleteness at high
redshifts, because any completeness correction requires the knowledge of
the luminosity function as a function of galaxy type, which is not determined accurately
enough, so we  investigate the local density contrast only at $z\leq 0.7$.

\subsection{Red and blue galaxies}\label{redbluedefinition}
From the COMBO-17 data we know the redshift and the SED for each galaxy, so it is possible to
calculate their absolute restframe magnitudes and colours. We can use this
information to investigate the properties of different galaxy types,
e.g. red sequence and blue cloud galaxies, where we use the
prescription of \citet{Bell04} to separate 
the two populations from each other:
\begin{equation}\label{redbluedef1}
{\rm Red\ sequence:\quad} (U-V) > (U-V)_{\mathrm {lim}}
\end{equation}
\begin{equation}\label{redbluedef2}
{\rm Blue\ cloud:\quad} (U-V) < (U-V)_{\mathrm {lim}} 
\end{equation}
\begin{equation}\label{redbluedef3}
(U-V)_{\mathrm {lim}} = 1.25-0.4 z -0.08(M_V-5\logten h+20)~,
\end{equation}
where $z$ denotes the redshift of a given galaxy. 

{\bf The observed bimodality in the colour-magnitude plane permits a
model-independent definition of the two different galaxy
populations. While the blue cloud consists of late-type, star-forming
galaxies, the well defined red sequence contains mainly early-type
quiescent galaxies.}

\section{Method}\label{Method}
The local density contrast
\begin{eqnarray}\label{overdensdef}
\delta=\frac{\rho-\overline{\rho}}{\overline{\rho}}~,
\end{eqnarray}
where $\rho$ is the local density, and $\overline{\rho}$ the mean
density, is commonly measured in spheres with a radius of 
e.g. $r=8\mpcoh$, as adopted by \citet{Croton05}.

Obviously, the redshift errors in our sample are too large to
calculate overdensities in such spheres, but it is still possible to
measure overdensities, albeit in slightly larger volumes. Instead of
counting the number of galaxies in a sphere or a cylinder centred on
individual galaxies, or distributed randomly within the survey volume,
we calculated the {\it comoving space densities\/} $\rho(z)$ in
redshift bins of $\Delta z = 0.02$ and steps of $\delta z
=0.005$.  At a redshift of $z=0.3$, this corresponds to to a
comoving radial bin size distance of $\Delta r\simeq 53 h^{-1}$\,Mpc.
The width of one COMBO-17
field at that redshift ($30'$) is approximately $7.4
h^{-1}$\,Mpc.

{\bf This measurement cannot be compared directly with those where
the overdensities are estimated in spheres ($r=8h^{-1}$\,Mpc for \citealp{Croton05}
or $r=5h^{-1}$\,Mpc for \citealp{Cucciati06}): in our case the comoving
distance along the line of sight is more than 7 times larger than the
transverse distance, which is of the order of magnitude of
the typical radius of the spheres. This geometry is enforced by the
photometric redshift errors, and means we measure trends averaged within
a larger volume. But at least this approach means we avoid uncertainties due to
small-scale peculiar velocities, which complicate the measurements in
smaller spheres. In the end, numerical simulations are required for a
full interpretation of any of these results.}

\subsection{Determination of the mean density}
The mean density $\overline{\rho}$ was estimated
in the range $0.25 \leq z \leq 0.7$:  This avoids
contamination from faint galaxies in the pre-selected cluster Abell 901 at
$z=0.165$, which could be scattered in redshift due to large
photometric redshift 
errors. Also, with a
total field size of $0.78$\,\sqdegr and the correspondingly small survey
volume at $z<0.25$, a measurement of the mean number density is always
dominated by cosmic variance. Redshifts larger than $z=0.7$ are
excluded, because the sample then
starts to become incomplete and noisy. 

\begin{figure}[h]
\centerline{\psfig{figure=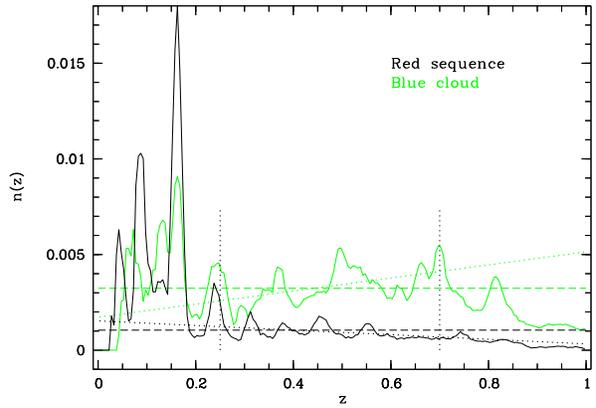,angle=270,clip=t,width=8.3cm}}
\caption[ ]{The mean comoving number density
of the three fields, for the blue (grey line) and red (black
line) subsamples. The dotted lines 
show the best fits (fitted in the range $0.25\leq z\leq 0.7$, {\bf as
indicated by the vertical dotted lines}) if an
evolution of the number density with redshift is assumed. The dashed
lines indicate the constant mean.
\label{meannumdens}}
\end{figure}

Fig.~\ref{meannumdens} shows the {\it mean} comoving number density
(the average of the three fields), for the blue and red subsamples. While
the number of red galaxies remains roughly constant with redshift, the
comoving number of blue galaxies tends to increase. We fit the trend (again
over the redshift range  $0.25 \leq z \leq 0.7$) with a straight line,
and use this empirical evolutionary fit instead of the constant mean in the calculation of
the overdensities, 
\begin{equation}
\bar{\rho}(z)=1.72\times
10^{-3}+3.42\times 10^{-3}\, z\; h^3{\rm Mpc}^{-3}.
\end{equation}
This fit certainly overpredicts the number density at $z\ge 0.8$, 
so we restrict this analysis to lower redshifts. We will mainly be
interested in practice in $z\simeq 0.3$, where the results have
little dependence on the strength of the assumed evolution.

\subsection{Redshift errors}\label{blurring}

Photometric redshift estimates have significant errors,
and we need to understand how these affect our estimates
of the density contrast. {\bf The comparison of our multicolour
redshifts with the available spectroscopic ones in the CDFS
(e.g. taken from the 2dFGRS and VVDS) showed no dependence of the
redshift errors on the galaxy type. With 17 filters the shapes of
the $SED$s are very well sampled and the 4000 \AA~break only plays a
minor role in the classification and redshift estimation of the
objects.  The 4000 \AA~ break only becomes a problem for faint galaxies ($R\ga
22$) at the
low- and high redshift end of the observations ($z\la 0.15$, and $z\ga
1.15$, respectively), when the break is
located at the edge of the filter set \citep{Wolf01b}.  However,} the
redshift accuracy depends on  
observed magnitude: statistically, blue galaxies are fainter than
red ones and thus tend to have slightly larger errors. Therefore, when
comparing the properties of red sequence with those of the blue 
cloud galaxies, we have to test which effect the different
error distributions have on the measured overdensities of our two subsamples. 

In order to simulate the influence of redshift errors on the
measurement of local densities in different volumes, we use a mock
galaxy survey based on a set of simulations by \citet{vanKampen99,vanKampen05}.
The phenomenological model predicts positions on the sky, 
redshifts including peculiar velocities,  magnitudes that would be
measured in the COMBO-17 filters, and absolute rest-frame 
luminosities and colours in the same bands that we use for the analysis
of the observations.

Four different simulation volumes are used to produce 80 different lightcones
representing individual COMBO-17 fields, for which we also calculate the
overdensities (Eq. \ref{overdensdef}). Each of these COMBO-17 mock samples is
selected in the same way as the observed data ($R\leq23.65$, $M_B - 5 \log_{10}h\leq-18.$).
Their number counts and overall redshift distribution have the 
same expectation, but they differ in
detail -- thus allowing us to assess the significance of `cosmic variance'.

For each galaxy in the COMBO-17 catalogue the rms error of its
estimated redshift is provided by the classification scheme, and we use
these errors to convolve the `spectroscopic' redshifts in our mock
sample with the error distribution. For each galaxy in the mock catalogue
we randomly pick out a value $\epsilon_z$ from the COMBO-17 data, 
then draw an error $\delta_z$ from a Gaussian distribution with
$\sigma=\epsilon_z$, and add this error to the given redshift. 
Using these new, `multicolour' redshifts, we can repeat the calculation
of the galaxy properties (e.g.  K-corrections and rest-frame
magnitudes), and of the overdensities. 

\begin{figure}[h]
\centerline{\psfig{figure=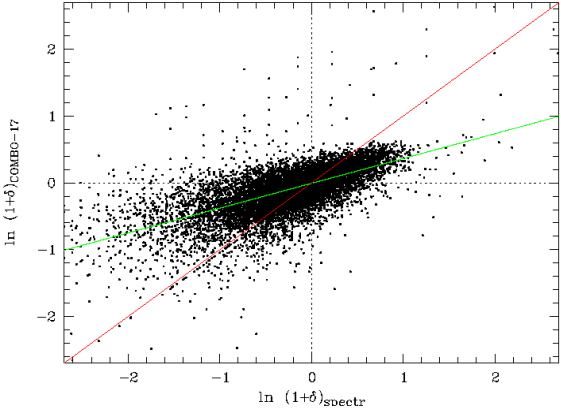,angle=270,clip=t,width=8.3cm}}
\caption[ ]{The overdensities in a mock COMBO-17 survey, calculated
for `spectroscopic' and `multicolour' redshifts in bins of $\Delta z =
0.02$ ($z\leq0.7$). A formal fit to the data points (dotted grey line) shows how the
amplitudes of both over- and underdensities are diminished.
\label{deltadelta}}
\end{figure}

Fig. \ref{deltadelta} shows the overdensities calculated
for `spectroscopic' redshifts against the `multicolour'
measurements in the same redshift bins. The scatter is {\bf small enough} to
facilitate a measurement of overdensities in a multicolour survey such as
COMBO-17. However, the tilt of the relation shows that high
overdensities become slightly lower, and deep underdensities slightly
shallower -- the dynamic range shrinks and the convolution with the
redshift error distribution washes out the measured structures.

In order to
facilitate a direct comparison between red sequence and blue cloud
galaxies, we have to understand what effect their slightly different 
redshift error distributions have on the measurement, and how we can
correct for any differences.  
We therefore now simulate redshift errors by drawing appropriate
{\it rms\/} errors from the red sequence and blue cloud
galaxy catalogues separately, and then calculate the 
colour-dependent overdensities for the mock sample. As can be seen from
Fig. \ref{erroranalysis}, in the presence of redshift inaccuracies
the existing small scale density fluctuations are washed out, and the amplitudes
of the overdensities are suppressed. Due to the statistically slightly larger errors of
the blue galaxies, the signal is more strongly suppressed than when the errors of
the red galaxies are applied. Thus, if we were to measure exactly the same structure
using a sample of red and blue galaxies as tracers,
the different redshift accuracies would cause us to infer a larger
overdensity (or smaller underdensity) from the red sample than from the
blue one. 

To account for this, we convolve the redshift distribution of the red
galaxies with a blurring function, which broadens their redshift {\it error}
distribution to make it resemble the redshift error
distribution of the blue galaxies. Of course the same procedure has to
be applied to the bright galaxies as well, in order to make them
comparable to the faint ones. In general, for each comparison we have to make
sure that the redshift distribution of the sample with the smaller
redshift errors has been blurred in order to make its error
distribution resemble the one with the lower accuracy.

The blurring function can be found via the convolution
theorem. Denote the redshift error by $\epsilon_z$, and let
$f$ and $g$ be the redshift error distributions of red and blue
galaxies respectively. We now seek a blurring function $b(\epsilon_z)$ that makes
them compatible:
\begin{eqnarray}\label{convolution}
f(\epsilon_z)=g(\epsilon_z)*b(\epsilon_z),
\end{eqnarray}
which is simply solved in Fourier terms: $B(k)=F(k)/G(k)$,
where $k$ is a wavenumber in redshift space.

\begin{figure}[h]
\centerline{\psfig{figure=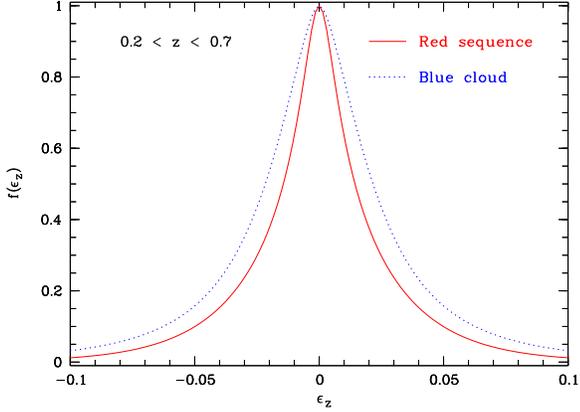,angle=270,clip=t,width=8.3cm}}
\caption[ ]{The redshift error distributions of red sequence (solid
  line) and blue
  cloud galaxies (dotted line), in the redshift range $0.2\leq z\leq
  0.7$. 
\label{errordistribplot}}
\end{figure}

\begin{figure}[h]
\centerline{\psfig{figure=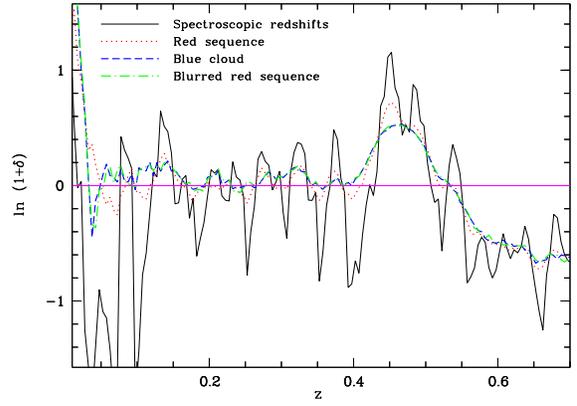,angle=270,clip=t,width=8.3cm}}
\caption[ ]{Overdensity of the mock galaxies in one `COMBO-17
field'. The solid line is the overdensity of the galaxies with
`spectroscopic' redshifts, the dotted line is the measurement using
the same galaxies, but convolved with the error distribution of the
red COMBO-17 galaxies, the dashed line is the measurement using the
blue COMBO-17 galaxies, and the dashed-dotted line is the overdensity
of the mock galaxies, which 
have first been convolved with the red error distribution, and then
further blurred in order to make their redshift inaccuracy comparable
to the ones that have been convolved with the blue error
distribution. 
\label{erroranalysis}}
\end{figure}

In order to evaluate the error distributions and account for its
redshift dependence, we calculate in each redshift bin of size
$\Delta z=0.1$ between $z=0$ and $z=1$ the sum of Gaussians {\bf where the
widths $\sigma_i$ are the rms errors of the redshift estimated by the
multicolour classification scheme \citep{Wolf01b,COMBOMain04}:}
\begin{eqnarray}
f(\epsilon_z)=\frac{1}{N_{\rm gal}}\sum_{i=1}^{N_{\rm gal}}{\frac{1}{2\pi\sigma_i^2}\exp\left(
 -\frac{\epsilon_z^2}{2\sigma_i^2}\right)}~,
\end{eqnarray}
for all colour and luminosity samples under consideration (for red
sequence and blue cloud galaxies see Fig. \ref{errordistribplot}). The
resulting functions can be closely approximated by a Breit-Wigner or Lorentz curve,
and it is convenient to treat this as the exact form:
\begin{eqnarray}
f(\epsilon_z) = \frac{W/2\pi}{\epsilon_z^2+\frac{W^2}{4}}~,
\end{eqnarray}
where $W$ is the full width at half maximum.
We parameterize the evolution with redshift (in the redshift range
$0.25\leq z \leq0.7$) of the full width at
half maximum $W$ with a second order polynomial:
\begin{eqnarray}
W(z)&=&a_0+a_1 (1+z)+ a_2 (1+z)^2~,
\end{eqnarray}
with the following coefficients for our red and blue, and bright ($M_\bj  - 5 \log_{10}h< -19.5$) and
faint ($M_\bj  - 5 \log_{10}h> -19.5$) subsamples.
Red sequence: $a_0=0.015$, $a_1=0.033$, $a_2=0.026$.
Blue cloud: $a_0=0.090$, $a_1=0.142$,
$a_2=0.068$. Bright galaxies:
$a_0=0.124$, $a_1=0.200$, $a_2=0.087$.
Faint galaxies: $a_0=0.094$, $a_1=0.164$, $a_2=0.082$.

The Fourier transform of a Lorentzian is
\begin{eqnarray}
F(k)=\exp\left(-W|k|/2\right)~,
\end{eqnarray}
from which is is readily seen that the required blurring function is
\begin{eqnarray}
b(\epsilon_z) = \frac{\Delta W/2\pi}{\epsilon_z^2+\frac{(\Delta W)^2}{4}}~,
\end{eqnarray}
where $\Delta W$ is the difference in widths of the two populations.
We can now use this probability distribution to degrade 
the redshift accuracy of a given sample in order to be
comparable to another sample with larger redshift errors: a redshift offset
is drawn randomly from the blurring probability distribution,
and added to the true data.

Fig. \ref{erroranalysis} also shows the overdensity of the mock galaxies, which
have first been convolved with the red error distribution, and then
further blurred in order to make their redshift inaccuracy comparable
to the ones that have been convolved with the blue error
distribution. Since the photometric redshift accuracies of the
sub-samples are now equal by construction, we can now start
to look for differences in the overdensity patterns as a function
of colour or luminosity.

\section{Results}\label{Results}
\subsection{Overdensities in the COMBO-17 survey}

\begin{figure}[h]
\centerline{\psfig{figure=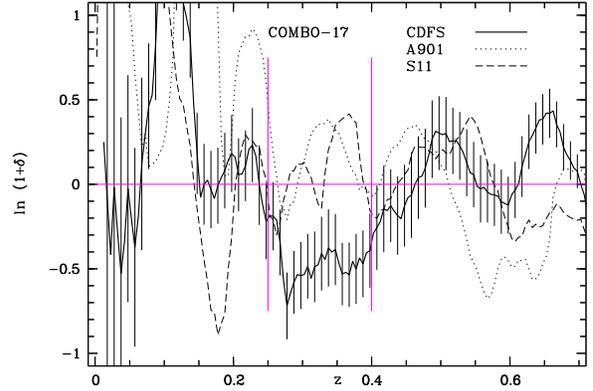,angle=270,clip=t,width=8.3cm}}
\caption[ ]{The overdensities in the three COMBO-17 fields, calculated
in bins of $\Delta z = 0.05$ (which corresponds to $\Delta r\simeq
132.5 h^{-1}$\,Mpc at $z=0.3$) and steps of $\delta z =0.01$, versus
redshift.  The vertical lines
indicate the range in which the luminosity function will be
estimated. The errors are rms errors calculated from 80 COMBO-17 mock
catalogues, but in order to avoid confusion error bars are only
plotted for every second data point. 
\label{deltaCOMBO}}
\end{figure}

Fig. \ref{deltaCOMBO} shows the overdensities measured in the three
COMBO-17 fields, which we calculated
in relatively large bins of $\Delta z = 0.05$ {\bf (which corresponds to $\Delta r\simeq
132.5 h^{-1}$\,Mpc at $z=0.3$),} in steps of $\delta z =
0.01$. Later we will decrease the size of our bins {\bf again}, but here we want
to compare the large-scale properties of the three fields. 

The error bars plotted for the CDFS are the variances of the
overdensities calculated in 80 mock COMBO-17 fields and should thus
include not only Poisson noise, but also cosmic variance. However,
since the data points are highly correlated (the spacing of the bins
being smaller than the binsize), it should be noted that the errors
are also correlated.

{\bf One well-known high-redshift structure in the CDFS, a sheet at $z=0.66$
(\citealp{Gilli03,COMBOMain04,Adami05}), clearly shows up in our
measurement (see Fig. \ref{deltaCOMBO}).} 
For the present paper, we are more interested in
the range $0.25 \leq z \leq 0.4$. Here one of our three
fields, the CDFS, is underdense with respect
to the others. The mean overdensity in the CDFS is
$\delta=-0.36\pm0.08$, whereas in the A901 field it is $\delta=
0.16\pm0.21$, and in the S11 field we find $\delta=0.11\pm0.24$,
respectively. So in both A901 and S11 the overdensity fluctuates about
the mean, whereas the CDFS is clearly underdense in this redshift range.

This is a fortunate coincidence: owing to the smaller number of
galaxies and the dynamic range of overdensities observed by COMBO-17 we can not split our sample  
into overdensity bins in the way e.g. \citet{Croton05} did -- 
but we can compare the statistical properties 
of the galaxies in this specific underdense region with those in `normal' dense
regions at the same redshift.

\begin{figure}[h]
\centerline{\psfig{figure=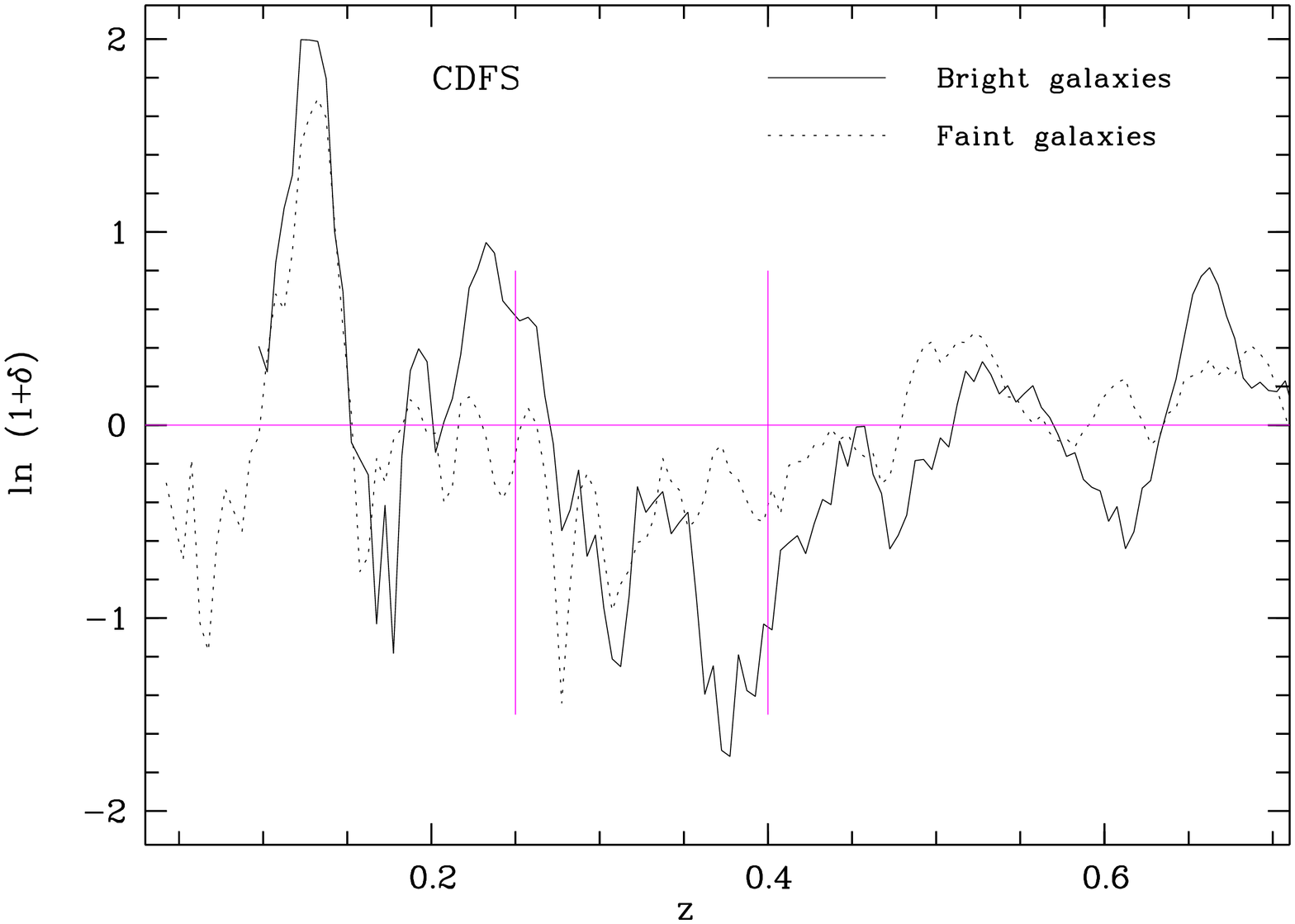,angle=270,clip=t,width=8.3cm}}
\centerline{\psfig{figure=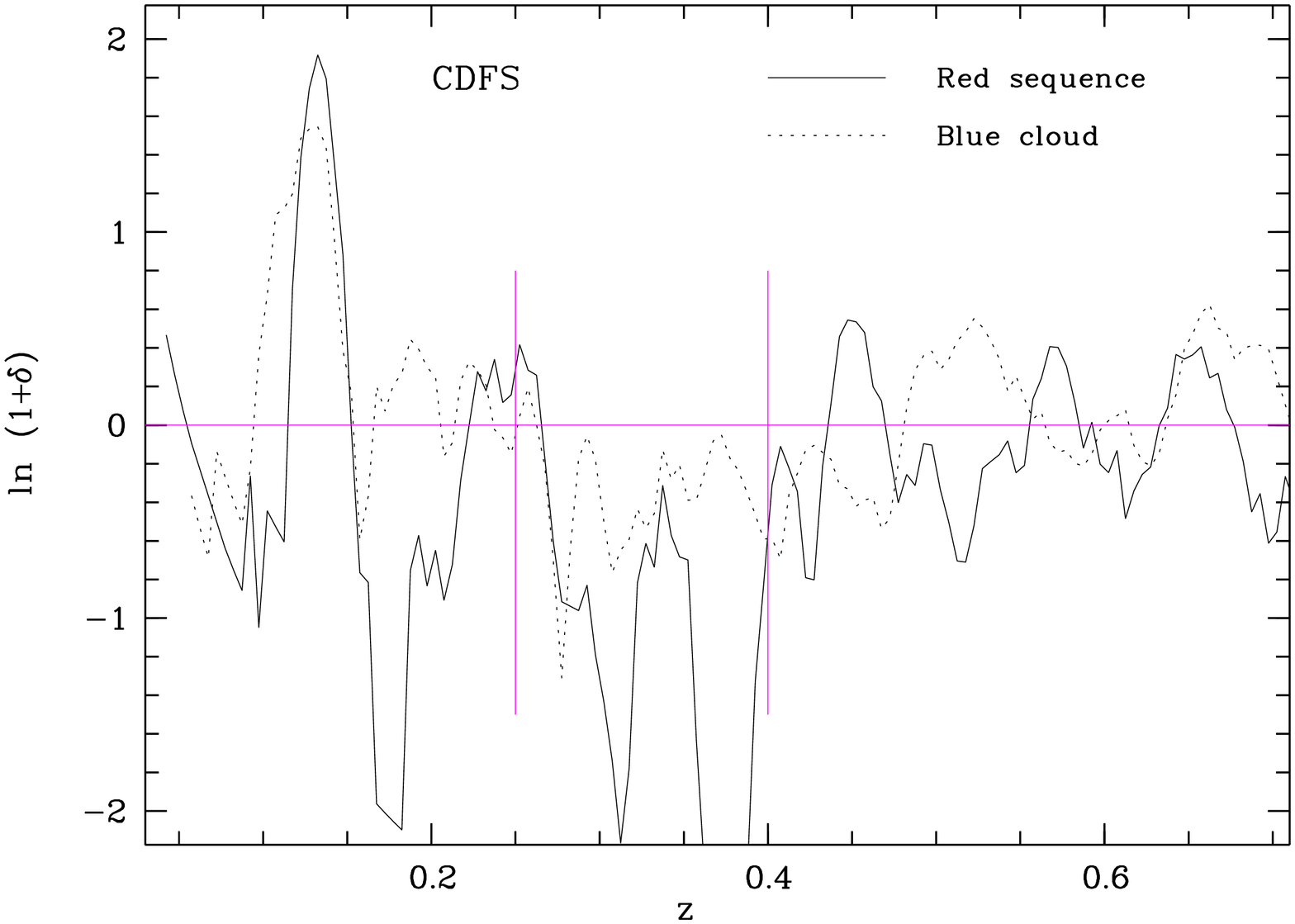,angle=270,clip=t,width=8.3cm}}
\caption[ ]{The overdensities in the CDFS, calculated in bins of $\Delta z = 0.02$, for bright ($M_\bj - 5 \log_{10}h< -19.5$), and faint
($M_\bj - 5 \log_{10}h\geq -19.5$) galaxies (upper panel), and red sequence and blue
cloud galaxies (lower panel). For both the bright and red sequence
subsamples, the redshifts have been blurred, in order to make their redshift
inaccuracies similar and thus the measurement of the overdensities
comparable to the faint and blue cloud 
subsamples, respectively.

\label{deltaCOMBOsubsamples}}
\end{figure}

Before measuring and comparing luminosity functions, we calculate
the overdensities in this field again 
for different subsamples (this time in smaller bins of size $\Delta z = 0.02$
and $\delta z=0.005$): a sample of red sequence and blue cloud
galaxies (see Sect. \ref{redbluedefinition}), and a sample of bright
($M_\bj  - 5 \log_{10}h< -19.5$), and faint ($M_\bj  - 5 \log_{10}h\geq -19.5$) galaxies,
respectively, see Fig. \ref{deltaCOMBOsubsamples}. The numbers of
galaxies in the different subsamples are given in Table \ref{subsampletab}

Unfortunately the mock COMBO-17 catalogues we used to calculate the
rms errors of the overdensities can currently not be used to calculate
errors for red and blue (or bright and faint) subsamples as well, since
 the mock galaxies do not exhibit the same
dependencies of colour and luminosity on the local density contrast as
the observed galaxies. A thorough error analysis has thus to be
postponed to a future paper, when improved mocks are available.

\begin{table}
\centering
\caption[ ]{The numbers of galaxies in the different subsamples,   
per COMBO-17 field. All galaxies are preselected to have $R\leq23.65$,
$M_\bj  - 5 \log_{10}h\leq-18.$, and $0.25\leq z \leq 0.4$. `Bright' means $M_\bj  - 5 \log_{10}h<
-19.5$, and `faint' $M_\bj  - 5 \log_{10}h\geq -19.5$\\\label{subsampletab}} 
\begin{tabular}{r|r r r r r  }
COMBO-17 field& $N_{\rm tot}$& $N_{\rm red}$& $N_{\rm blue}$&$N_{\rm bright}$&$N_{\rm faint}$\\ \hline
CDFS&$301$&$56$&$245$&$63$&$238$\\
A901&$594$&$162$&$432$&$165$&$429$\\
S11&$543$&$178$&$365$&$141$&$402$\\
\end{tabular}
\end{table}

From Fig.~\ref{deltaCOMBOsubsamples}, it is evident that although the redshifts of the red and
bright samples have been further smoothed in 
the way explained in Section \ref{Method}, the structures are  more distinct in
the red/bright samples than in the blue/faint ones, respectively.

The different samples trace the underlying dark matter density field
differently. Bright  galaxies are
generally found to be  more strongly clustered than the faint 
ones, because they
are thought to reside in  massive dark matter haloes, which are
generally believed to be more strongly clustered than small ones (e.g. 
\citealp{ColeKaiser89,Mo96,ShethTormen99}). At the same time,  red
galaxies are observed to be more strongly clustered than the blue
galaxies
(e.g. \citealp{Davis76,Norberg02,Zehavi05,PhlepsMeise03,Phleps06,Meneux06}). 
However, it is presently
not clear whether luminosity or colour is the determining property
(see e.g. \citealp{Norberg02}).

As can be seen in Fig. \ref{deltaCOMBOsubsamples}, the underdensity in
the CDFS at $0.25\leq z \leq 0.4$ is particularly pronounced when
calculated using only red galaxies for the determination -- this
region is mainly deficient 
in red galaxies. We will see in the next section that this deficiency
reflects mainly a reduction in the number of {\it faint\/} red galaxies.

\subsection{Luminosity functions}

In order to investigate which  galaxies  are most deficient
in the underdense region in the CDFS, we 
we have calculated rest-frame $B$-band luminosity functions for the
galaxies in the redshift bin $0.25\leq z \leq 0.40$ in all three
fields, split by colour according to Eqn. \ref{redbluedef3}.

At redshift $z=0.4$, a luminosity of $M_B - 5 \log_{10}h=-17$ corresponds to
an observed-frame apparent magnitude of $R_{\rm tot}=23.2$ or $R_{\rm
aper}=23.5$ in the COMBO-17 apertures. The aperture magnitudes and
colours determine the completeness, which we estimate as $>90$\% at
every point in the redshift-luminosity data cube. Nevertheless, the
calculation of the luminosity functions has been implemented exactly as described in W03
 and later
COMBO-17 papers, {\bf where the non-parametric $1/V_{\rm max}$ estimator
  \citep{Schmidt68} is used in the form proposed by
  \citet{DavisHuchra82} and modified by \citet{Fried01}, who for the
  first time took completeness
  corrections into account in the calculations. Any
  redshift bins where the magnitude cutoff of the survey shrinks
  the accessible volume of the bin by more that 30\% compared to an
  infinitely deep survey are ignored, thus ensuring that the faint
  end of the luminosity function is correctly represented.}

\begin{figure*}[h]
\centerline{\psfig{figure=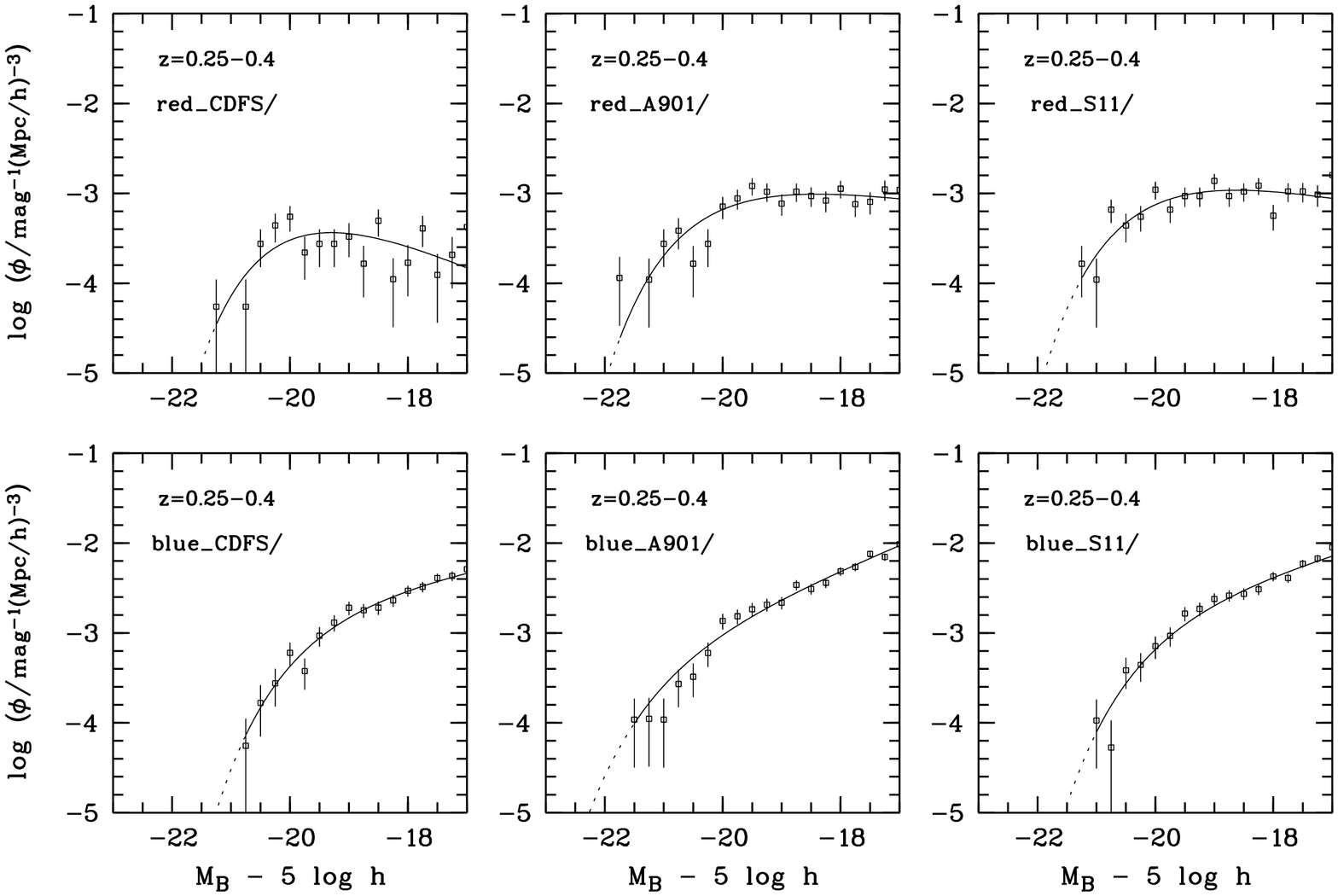,angle=270,clip=t,width=19.cm}}
\caption[ ]{The luminosity function of the red sequence (upper panel)
and blue cloud (lower panel) galaxies in
the redshift range $0.25 \leq z \leq 0.4$,
 for the three COMBO-17 fields.
The STY fit is overplotted in each panel.\label{lumifunc}}
\end{figure*}

Fig. \ref{lumifunc} shows the
luminosity functions of the red sequence and blue cloud galaxies in the
redshift bin $0.25\leq z \leq 0.4$
for  the three COMBO-17 fields.
Parameters for the Schechter fits (as plotted in Fig. \ref{lumifunc}) are given
in Table \ref{STYparameters}. 
We present the luminosity functions separately for our three fields in
order to investigate their differences. In W03 it was already reported
that the CDFS is underdense in the `semi-local' redshift bin
$z=[0.2,0.4]$ (see their Fig. 12). However, W03 investigated luminosity functions either
split by field or split by spectral type. In contrast, here we present
the LF split both by field and by rest-frame colour.

\begin{table*}
\centering
\caption[ ]{STY fit parameters for the luminosity functions of red
  sequence and  blue cloud galaxies {\bf in the redshift range
  $0.2\leq z\leq0.4$. The numbers of the galaxies per subsample and
  field are given in Table \ref{subsampletab}.}\\\label{STYparameters}} 
\begin{tabular}{llcrc}
sample& field  & $M^*-5$~log~$h$ (Vega mag) & $\phi^* \times 10^{-4} h^{3}{\rm Mpc}^{-3}$ & 
  $\alpha$ \\
\noalign{\smallskip} \hline \noalign{\smallskip}
&CDFS        & $-19.53\pm 0.33$ & $ 25.00\pm  1.39$ & $-1.34\pm 0.17$ \\
blue&A901    & $-20.88\pm 0.38$ & $  9.23\pm  0.39$ & $-1.68\pm 0.09$ \\
&S11         & $-19.89\pm 0.31$ & $ 22.65\pm  1.06$ & $-1.49\pm 0.12$ \\
\noalign{\smallskip} \hline

&CDFS        & $-19.64\pm 0.41$ & $ 10.35\pm  1.25$ & $-0.27\pm 0.35$ \\
red&A901     & $-20.10\pm 0.31$ & $ 18.19\pm  1.31$ & $-0.79\pm 0.17$ \\
&S11         & $-19.97\pm 0.28$ & $ 22.49\pm  1.54$ & $-0.71\pm 0.17$ \\
\noalign{\smallskip} \hline
\end{tabular}
\end{table*}

As can be seen from Fig. \ref{lumifunc} and Table \ref{STYparameters},
the luminosity function of the blue cloud galaxies does not differ
from field to field (apart from the normalisation $\phi^*$, which 
unsurprisingly is
lower in a low density region). In contrast, the luminosity
function of the red sequence galaxies in the CDFS (the underdense
region) is indeed clearly distinct from the one measured in the two other
fields, which have about mean density. Not only is the normalisation
slightly lower, but also the slope $\alpha$ is clearly more positive:
the underdensity in the CDFS is mainly due to a deficiency of faint
red galaxies.

\section{Discussion}\label{discussion}

Our detection of a lack of faint red galaxies in voids
is in qualitative agreement with the work of \citet{Croton05}, who
investigated the influence of the environment on galaxy properties in
the local universe {\bf ($0.05\leq z\leq0.15$)} using 
2dFGRS data \citep{Colless01,Colless03}. Croton et al. were able to measure the
type-dependent luminosity functions in six different overdensity
regimes from  voids to clusters of galaxies. They found that late-type
galaxies display a consistent luminosity function across all density
environments, with a weak dimming of $M^*$ in the underdense regions
and an almost constant faint-end slope. In contrast the luminosity
function of the red galaxies differs sharply between the extremes in
environment: $M^*$ brightens by approximately 1.5 mag going from voids
to clusters, while the faint-end slope moves from $\alpha \simeq
-0.3$ in underdense regions to $\alpha \simeq
-1.0$ in the densest part of the survey.

A similar analysis has been undertaken by
\citet{Zandivarez06}, who investigated the variation of the galaxy
luminosity function {\bf at $0.02 \leq z\leq 0.22$} with the mass of galaxy groups identified in
the Fourth Data Release of the Sloan 
Digital Sky Survey \citep{Adelman06}, and found a continuous
brightening of the characteristic magnitude, and a steepening of the
faint end slope as the group mass increases. When they split their sample
by $u-r$ colour into red and blue galaxies, they found
that the changes observed as a function of group mass are mainly
seen in the red, passively evolving, galaxy population, while the
luminosities of blue galaxies remain almost unchanged with mass. When
we take the group mass as correlating with the local density, then this
result is consistent with the result of both \citet{Croton05}, and
with our own.

Therefore we conclude that the same dependency of the luminosity
function on environment -- a lack of faint red galaxies in underdense
regions and a dominant population of bright red galaxies in overdense
environments -- was already in place up to $z\simeq 0.4$.
We now have to ask what is known about environmental trends at higher redshifts.

Results from the DEEP2 Galaxy Redshift Survey \citep{Davis03} show
that the colour segregation observed between local group and field
galaxies is even seen at $z\sim 1$
\citep{Cooper06a,Cooper06b,Gerke06}. DEEP2 is a spectroscopic
survey of galaxies at redshifts around unity ($0.7 \la z \la 1.4$), to
a limiting magnitude of $R_{AB}=24.1$. The unprecedented combination
of depth and redshift accuracy allows for an examination of the
influence of the environment on the galaxies' properties at $z\sim
1$. \citet{Cooper06b} use a sample of $19\,464$ galaxies drawn
from the DEEP2 survey to show that the colour-density relation evolves
continuously, with red galaxies more strongly favouring overdense
regions at lower redshift as compared to their high-redshift
counterparts, with the fraction of blue galaxies (which is lower in
groups than in the field) staying roughly constant with
redshift. However, at $z\simeq 1.3$, the red fraction starts to
correlate only weakly with overdensity \citep{Cooper06a}, and the
group and field blue fractions become indistinguishable
\citep{Cooper06b}.

{\bf \citet{Cucciati06} also carried out an investigation of the redshift
and luminosity evolution of the colour-density relation using data
from the VVDS \citep{LeFevre05}, and also found that the trend for red(/blue) galaxies to be found
mainly in dense(/underdense) regions seen at lower redshifts
progressively disappears in the highest redshift bins investigated
($1\la z \leq 1.5$).}

{\bf \citet{Ilbert06a} reconstructed the 3D density field using a
  Gaussian filter of smoothing length $5h^{-1}$\,Mpc, and estimated
  the luminosity function of 6582 galaxies of the VVDS in four
  redshift bins between $z=0.25$ and $z=1.5$, for galaxies in
  overdense and underdense environments, respectively. They find a
  strong dependence of the luminosity function on environment up to
  $z=1.2$,  that is, a steeper slope in underdense regions, and a
  steepening with increasing redshift. In the
  redshift range $0.6\leq z \leq 0.9$  they split the sample into red
  and blue galaxies, and again find the slope to be steeper in
  underdense regions, independent of spectral type. This is different
  from our results (where the slope of the luminosity function of blue
  galaxies remains unaffected by the local density, whereas for the
  red galaxies it changes), but their {\it interpretation}, albeit with a slightly
  different perception of the results, is compatible
  with ours: Together with the
  observations of \citet{Cucciati06}, they interpret their result not
  as a lack of faint red galaxies in 
  underdense regions, but as an increase of their number density in
  overdense regions with cosmic time. 
}

This generally observed trend -- the growing fraction of red galaxies in
overdense regions, while the overall fraction of blue galaxies evolves slowly
up to $z\simeq 1$ -- suggests that the strong dependence
of the galaxy properties on the environment found  at lower redshifts is a
result of environment-driven mechanisms. The build-up of the red
sequence appears to have occurred preferentially in overdense regions.
One further piece of evidence in this direction come from
\citet{Gerke06}, who find that at $z\geq 0.7$ red galaxies
already tend to be bright, and bright galaxies in general tend to live in
dense environments, even at redshifts around unity.

Our results complement and reinforce this general picture, and fit
into the standard model of hierarchical clustering 
growth and galaxy evolution: bright, massive  galaxies formed early in the
rare, highly clustered high-mass peaks of the dark matter
distribution \citep{Kaiser84} and are thus more strongly clustered
than faint, less massive galaxies which have formed later in less
clustered environments. But not only are
bright galaxies more strongly clustered than faint ones, red
galaxies are in addition much more strongly clustered than blue galaxies. It is
commonly believed that the red, early-type 
galaxy population are remnants of merger processes, whereas the
blue galaxies form stars at a rate only determined by their internal
physical properties (\citealp{Baldry04,Bell04}), independent of their
environment. In low-density regions, galaxies typically reside in the centres of
low-mass dark matter haloes and are thus faint. Since there is still
gas available for star formation, they are also blue. The
merger rate is low, so they are mainly spirals. In higher density
regions the typical galaxy is a central galaxy of a more massive dark
matter halo, so it is tends to be bright. There is no gas left for
star formation, and the merger rate was high, so such galaxies are rather
red and early type. 
This interpretation is supported by the fact that the luminosity function of blue galaxies
remains unchanged with the local density contrast, whereas the luminosity function of
the red galaxies depends on the environment.

\section{Summary and outlook}\label{Summary}

The dependence of galaxy properties on the environment in which they reside is
a clue to the physical processes that led to
their formation and present appearance: {\bf If the local density contrast changes the
path the evolution of a galaxy takes (by merging, gas stripping, etc.), then this should be reflected in
the properties of the galaxies that inhabit different
environments.} The means of investigating 
this correlation of galaxy properties and the local density contrast
is the type- {\bf or colour-}dependent luminosity function, calculated in different
density regimes.

 These measures, local overdensities and luminosity
functions, make different demands on the data: For a precise
determination of overdensities good redshift quality is needed, in the
most optimal case spectroscopic. But current spectroscopic redshift
surveys are {\bf either not deep enough or have too small statistics} to
allow for a precise measurement of the 
luminosity function, especially at intermediate to high redshifts.

In this paper, we have demonstrated {\bf
  how} multicolour surveys can {\bf be used to}
overcome this problem. {\bf Multicolour surveys have larger redshift
  errors than spectroscopic ones, and redshift inaccuracies smooth out the
structures. The extent to which the amplitudes are suppressed depends on the size of the
redshift errors, but if the redshift {\it error} distribution is well
  determined, this can be taken into account.

Red galaxies have -- statistically -- better redshifts than blue
galaxies (because the redshift accuracy depends on magnitude, and red
galaxies are statistically brighter than blue ones). So when we intend
to compare their overdensities, we have to blur the good red
galaxies' redshifts in order to make their redshift error distribution
resemble that of the blue ones.  The method, which makes use of the
  convolution theorem, was successfully tested
  with a mock COMBO-17 survey.}

We have used the COMBO-17 survey to calculate overdensities for different
samples of galaxies (a red, blue, bright and faint subsample,
respectively), in three fields. {\bf In order to make the measurements
comparable to each other, the redshifts of the subsample possessing
the smaller redshift errors were blurred before calculating the overdensities.}
Instead of calculating the overdensity
in small spheres, as is usually done, we do it in thin redshift
slices. We find
that {\bf one of the three COMBO-17 fields}, the
Chandra Deep Field South (CDFS), displays a relatively large underdense 
region, where the other two fields have overdensities fluctuating
about mean density. We use this to compare the luminosity functions
of red and blue galaxies in different density regimes (but at the same
redshift, $0.25 \leq z \leq 0.4$). 

The luminosity function of the blue cloud galaxies is unaffected by
the environment: it has the same shape in all three fields. The
luminosity function of the red sequence galaxies, on the other hand,
is very different in the underdense region in the CDFS: its faint-end slope
$\alpha$ is significantly more positive than in the other two fields
at the same redshifts. 
This finding -- that the underdensity is mainly due to a lack of faint
red galaxies -- is consistent with results at lower redshift
(e.g. \citealp{Croton05} or \citealp{Zandivarez06}), and fits into the
common picture of hierarchical galaxy formation.

Our present analysis is only a preliminary study of how multicolour
data can be used to investigate the dependence of galaxy properties on
the local environment at redshifts $z\ga 0.2$. A full quantification
of the effect of the environment on 
galaxy properties will require much larger surveys. First of all the survey volumes have to 
be larger: not only will the statistics be better in a bigger survey,
but also the dynamic range of observed overdensities. In COMBO-17,
the range of overdensities that can be investigated is limited. In a
large-area survey, the field can be split into many different smaller
subfields (either randomly distributed or deliberately chosen by surface density)
and a similar analysis to ours can be carried out, or a 
count-in-(large)-cells analysis similar to the one by \citet{Wild05} and
\citet{Conway05}, where they counted galaxies in approximately cubical
boxes.

Second, a completeness to fainter
magnitudes is desirable for a correct and precise
determination of the slope $\alpha$ of the luminosity
function also at higher redshifts. This is  important for the
investigation of the {\it evolution} of the dependence of galaxy
properties on 
environment.

We can look forward to achieving many of these goals
with new generations of deep multicolour or photometric redshift
surveys, such as VST-16, KIDS \citep{Kuijken06}  
or Pan-STARRS \citep{Kaiser05}.

\begin{acknowledgements}
S. Phleps acknowledges financial support
by the SISCO Network provided through the European
Community's Human Potential Programme under contract
HPRN-CT-2002-00316.
JAP was supported by a PPARC Senior Research Fellowship.
CW was supported by a PPARC Advanced Fellowship.
We would like to express our appreciation of the helpful discussions
we had with Peter Schuecker, who sadly died in November 2006.

\end{acknowledgements}

\bibliographystyle{aa}

\end{document}